\RequirePackage{fix-cm}
\documentclass[smallcondensed]{svjour3}     
\smartqed  
\usepackage{graphicx}
\usepackage{amsmath}
\usepackage{amssymb}
\journalname{myjournal}
\begin{document}

\title{On the Cross-Correlation of a $p$-ary m-Sequence and its Decimated Sequences by $d=\frac{p^n+1}{p^k+1}+\frac{p^n-1}{2}$
}


\author{Sung-Tai Choi \and Ji-Youp Kim \and \\Jong-Seon No}


\institute{ The material in this paper will be very partly presented at the 2012 IEEE International Symposium on Information Theory.\\
S.-T. Choi \and J.-Y. Kim \and J.-S. No \at
              Department of Electrical Engineering and
Computer Science, INMC\\ Seoul National University, Seoul 151-744, Korea\\
              \email{stchoi@ccl.snu.ac.kr}           
           \and
           J.-Y. Kim \at
              \email{lakroforce@ccl.snu.ac.kr}
            \and  
               J.-S. No \at
              \email{jsno@snu.ac.kr}
}

\date{Received: date / Accepted: date}

\maketitle

\begin{abstract}
In this paper, for an odd prime $p$ such that $p\equiv 3\bmod 4$, odd $n$, and $d=(p^n+1)/(p^k+1)+(p^n-1)/2$ with $k|n$, the value distribution of the exponential sum $S(a,b)$  is calculated as $a$ and $b$ run through $\mathbb{F}_{p^n}$.  The sequence family $\mathcal{G}$ in which each sequence has the period of $N=p^n-1$ is also constructed. The family size of $\mathcal{G}$ is $p^n$ and the correlation magnitude is roughly upper bounded by $(p^k+1)\sqrt{N}/2$.  
The weight distribution of the relevant cyclic code $\mathcal{C}$ over $\mathbb{F}_p$ with the length $N$ and the dimension ${\rm dim}_{\mathbb{F}_p}\mathcal{C}=2n$  is also derived. Our result includes the case in \cite{Xia} as a special case.
\end{abstract}


\section{Introduction}
\label{Intro}
There have been lots of researches on the cross-correlation between decimated m-sequences. Let $p$ be an odd prime and $\mathbb{F}_{p^n}$ the finite field with $p^n$ elements. The cross-correlation function corresponds to the exponential sum given as
\begin{align}\label{Exp_Sum}
	S(a,b)=\sum_{x\in \mathbb{F}_{p^n}}\chi\big(ax^{d_1}+bx^{d_2}),~a,b\in \mathbb{F}_{p^n} 
\end{align}
where ${\rm tr}_1^n(\cdot)$ is the trace function from $\mathbb{F}_{p^n}$ to $\mathbb{F}_p$, $\chi(\cdot)=\omega^{\rm tr_1^n(\cdot)}$ is a canonical additive character of $\mathbb{F}_{p^n}$, and $\omega=e^{2\pi\sqrt{-1}/p}$ is a primitive $p$-th root of unity.

The research on $S(a,b)$ can be exploited as: 
  
\noindent 1) When the magnitudes of $S(a,b)$ is small enough, it can be used to construct a new sequence family with low correlation property  \cite{Gold}--\cite{Kim2};\\
\noindent 2) Value distribution of $S(a,b)$ is used for the calculation of the weight distribution of the corresponding cyclic codes \cite{Feng}--\cite{Luo3}.

If the number of distinct values of $S(a,b)$ when $a$ and $b$ run through $\mathbb{F}_{p^n}$ is small enough, then the value distribution of  $S(a,b)$ is likely to be derived. However, although its value distribution is hard to derive due to some technical problems, the upper bound on the magnitudes of $S(a,b)$ is still meaningful for the construction of sequence families with low correlation property. The general methodology to drive the value distribution of $S(a,b)$ is formulated in \cite{Choi}.

 Not much is known about the weight distributions of cyclic codes except for very specific cases. Especially,  for the alphabet size of an odd prime $p$, the value distribution of $S(a,b)$ and the weight distribution of the corresponding cyclic code were derived for $d_1=1$ and $d_2=(p^k+1)/2$ in \cite{Luo}. In \cite{Luo2}, the same work is done for $d_1=2$ and $d_2=p^k+1$. Recently, the value distribution of $S(a,b)$ for $d_1=1$, $d_2=(p^n+1)/(p+1)+(p^n-1)/2$, odd prime $p$ such that $p\equiv 3\bmod 4$, and odd $n$ is derived in \cite{Xia}.

In this paper, the value distribution of $S(a,b)$ is calculated for $d_1=1$ and  $d_2=(p^n+1)/(p^k+1)+(p^n-1)/2$ with $k|n$, an odd prime $p$ such that $p\equiv 3\bmod 4$, and odd $n$.  Using the result, the maximum magnitude of cross-correlation values of the sequence family $\mathcal{G}$ and the weight distribution of the cyclic code $\mathcal{C}$ are derived, respectively. Our result includes the result in \cite{Xia} as a special case.  

This paper is organized as follows. In Section \ref{Sec_Not}, preliminaries are stated. In Section \ref{Sec_ValDis_Sab}, the value distribution of $S(a,b)$ is derived. In Section \ref{Sec_Seq_Fam}, the upper bound of cross-correlation magnitude of the sequence family $\mathcal{G}$ is calculated. In Section \ref{WeightDist_C}, the weight distribution of the cyclic code $\mathcal{C}$ is obtained. The conclusion is given in Section \ref{Sec_Con}. 

\section{Preliminaries}
\label{Sec_Not}

\subsection{Exponential Sum $S(a,b)$ and the Hamming Weight of the Code $\mathcal{C}$}

Let $p$ be a prime and $\mathbb{F}_{p^n}$ the finite field with $p^n$ elements. Then the trace function 
${\rm tr}_k^n(\cdot)$ from $\mathbb{F}_{p^n}$ to $\mathbb{F}_{p^k}$ is defined as
\begin{align*}
	{\rm tr}_k^n(x)=\sum_{i=0}^{\frac{n}{k}-1}x^{p^{ki}}
\end{align*}
where $x\in \mathbb{F}_{p^n}$ and $k|n$. Let $\alpha$ be a primitive element of $\mathbb{F}_{p^n}$ and $\mathbb{F}_{p^n}^*=\mathbb{F}_{p^n}\setminus \{0\}$. 

We will consider
\begin{align}\label{Def_Exp_Sum}
	S(a,b)=\sum_{x\in \mathbb{F}_{p^n}}\chi (ax+bx^d), 
\end{align}
which is the case when $d_1=1$ in (\ref{Exp_Sum}). 

Let $\mathcal {C}$ be the cyclic code over $\mathbb{F}_p$ with the length $N=p^n-1$, in which each codeword is defined as
\begin{align*}
	c(a,b)=(c_0,c_1,\dots,c_{N-1}),~a,b\in \mathbb{F}_{p^n}
\end{align*} 
where $c_i={\rm tr}_1^n (a\alpha^i+b\alpha^{di})$, $0\leq i \leq N-1$. The Hamming weight of the codeword $c(a,b)$ is defined as
\begin{align*}
	H_w(c(a,b))&=\big|\{i|0\leq i\leq N-1,c_i\neq 0\}\big|.
\end{align*}

\subsection{Quadratic Form}

We define a quadratic form in $e$ variables over $\mathbb{F}_{p^k}$ as a homogeneous polynomial in $\mathbb{F}_{p^k}[x_1,\dots,x_e]$
\begin{equation}
	f({\bf x})=f(x_1,\dots,x_e)=\sum_{i,j=1}^e a_{ij}x_ix_j\nonumber
\end{equation}
where $p$ is an odd prime and $a_{ij}=a_{ji}\in \mathbb{F}_{p^k}$. We then associate $f$ with the $e\times e$ symmetric matrix $A$ whose $(i,j)$ entry is $a_{ij}$. The matrix $A$ is called the coefficient matrix of $f$ and $r$ denotes the rank of $A$. Then, there exists a nonsingular $e\times e$ matrix $B$ over $\mathbb{F}_{p^k}$ such that $H=BAB^T$ is a diagonal matrix, that is,  $H={\rm diag}(h_1,\dots,h_r,0,\cdots,0)$, where $h_i\in \mathbb{F}_{p^k}^*$. Let $\Delta=h_1\cdots h_r$, which will be used in the following lemmas.

A quadratic form $f(\bf x)$ in $e$ variables over $\mathbb{F}_{p^k}$ can
be regarded as a mapping $f(x)$ from $\mathbb{F}_{p^{ek}}$ to $\mathbb{F}_{p^k}$, when $x_i\in \mathbb{F}_{p^k}$. Thus,
we will also use the term `quadratic form' for this mapping $f(x)$
in $\mathbb{F}_{p^{ek}}$.  

If $f$ is a quadratic form over $\mathbb{F}_{p^k}$ and $b\in \mathbb{F}_{p^k}$, then an explicit formula for the number of solutions of the equation $f(x_1,\dots,x_e)=b$ in $(\mathbb{F}_{p^k})^e\approx \mathbb{F}_{p^{ek}}$ can be given. Hence the `quadratic form' can be exploited to evaluate $S(a,b)$ if the $S(a,b)$ is represented as a quadratic form. 
In the remainder of this section, some useful lemmas on a quadratic form are listed as follows.

\begin{lemma}\label{Lem_Cor_Qf}
Consider the function given by
\begin{equation}
	{\rm tr}_1^{n}\Big(\sum_i a_i x^{p^{ki}+1}\Big)={\rm tr}_1^k\big(f(x)\big)\nonumber
\end{equation}
where $i\geq 0$ are integers, $a_i\in \mathbb{F}_{p^n}^*$, and $k|n$. Then 
\begin{equation}\label{Eq_0516_A}
f(x)={\rm tr}_k^n\Big(\sum_i a_i x^{p^{ki}+1}\Big)
\end{equation} 
is a quadratic form in $n/k$ variables over $\mathbb{F}_{p^k}$. 
\end{lemma}
\begin{proof}
Any $x\in \mathbb{F}_{p^n}$ is represented as
\begin{align}\label{Eq_0516_B}
	x=x_1\alpha_1+x_2\alpha_2+\cdots x_e\alpha_e,~x_i\in \mathbb{F}_{p^k}
\end{align}
where $e=n/k$ and $(\alpha_1,\alpha_2,\dots,\alpha_e)$ is a basis of $\mathbb{F}_{p^n}$ over $\mathbb{F}_{p^k}$. Substituting (\ref{Eq_0516_B}) into (\ref{Eq_0516_A}), we have
\begin{align*}
	f(x)&={\rm tr}_k^n\Big(\sum_{i}a_i \big(\sum_{j=1}^e x_j\alpha_j^{p^{ki}}\big)\big(\sum_{l=1}^e x_l\alpha_l \big)\Big)\\
	&= \sum_{j=1}^e\sum_{l=1}^e x_j x_l {\rm tr}_k^n\Big(\alpha_l \sum_i a_i \alpha_j^{p^{ki}}\Big),
\end{align*}
which is a quadratic form with $e$ variables over $\mathbb{F}_{p^k}$.  \hfill $\Box$
\end{proof}

\vspace{0.1cm}

\begin{lemma}[Luo and Feng \cite{Luo}]\label{Lem_Cal_Rank}
The rank $r$ of the quadratic form $f(x)$
from $\mathbb{F}_{p^{ek}}$ to $\mathbb{F}_{p^k}$ is determined from the number of
elements that the form is independent of, i.e., $(p^k)^{e-r}$ is
the number of $y\in \mathbb{F}_{p^{ek}}$ such that $f(x+y)-f(x)-f(y)=0$ for all $x\in
\mathbb{F}_{p^{ek}}$.
\end{lemma}

\vspace{0.1cm}

\begin{lemma}[Luo and Feng \cite{Luo}]
\label{Lema_QfEvaluation}

Let $f(x)$ be a mapping from $\mathbb{F}_{p^{ek}}$ to $\mathbb{F}_{p^k}$ corresponding to the
quadratic form $f({\bf x})\in \mathbb{F}_{p^k}[x_1,x_2,\dots,x_e]$ of rank $r$
with $\Delta$. Then we have
\begin{eqnarray}
	\sum_{x \in \mathbb{F}_{p^{ek}}} \omega^{{\rm tr}^{k}_1(f(x))}=
	\begin{cases}
		\eta (\Delta) (p^k)^{e-\frac{r}{2}},&~{\rm if}~p^k\equiv 1\bmod 4\\
		j^r \eta(\Delta)(p^k)^{e-\frac{r}{2}},&~{\rm if}~p^k\equiv 3\bmod 4
	\end{cases}
\end{eqnarray}
where $j=\sqrt{-1}$ and $\eta(\cdot)$ is the quadratic character of $\mathbb{F}_{p^k}^*$ defined as
\begin{align*}
	\eta(x)=
	\begin{cases}
		1,&~{\rm if}~x~{\rm is~a~square~in~}\mathbb{F}_{p^k}^*\\
		-1,&~{\rm if}~x~{\rm is~a~nonsquare~in~}\mathbb{F}_{p^k}^*.
	\end{cases}
\end{align*} 
\end{lemma}

\subsection{Linearized Polynomial}

Let $q$ be a power of prime. A polynomial of the form
\begin{equation}\label{Eq_LinPoly}
\phi(x)=\sum_{i}a_ix^{q^{i}}\nonumber,
\end{equation}
where $a_i\in \mathbb{F}_{q^m}$, is called a linearized polynomial
over $\mathbb{F}_{q^m}$. If $F$ is an arbitrary extension field of $\mathbb{F}_{q^m}$ which includes the roots of $\phi(x)$,
then
\begin{eqnarray*}
\phi(\beta+\gamma)&=&\phi(\beta)~+\phi(\gamma),~~{\rm for ~all}~\beta,\gamma\in F\\
\phi(c\beta)&=&c\phi(\beta),~~~~~~~~~~~{\rm for ~all}~\beta\in F~{\rm and}~c\in \mathbb{F}_{q}.
\end{eqnarray*}
Hence the set of solutions of $\phi(x)=0$ in $F$ forms a vector
subspace over $\mathbb{F}_{q}$, i.e., the number of solutions in $F$ of $\phi(x)=0$ is equal to a power of $q$.

 \subsection{Weil's Bound} 
 
 The following lemma provides the upper bound on the magnitudes of the exponential sums, which is known as Weil's bound. 
\begin{lemma}[Theorem 5.38 \cite{Lidl}]
	\label{Thrm_Weil}
    Let $f(x)\in \mathbb{F}_{p^n}[x]$ be a polynomial of degree $l\geq 1$ with $\gcd(l,p^n)=1$ and let $\chi$ be a nontrivial additive character of $\mathbb{F}_{p^n}$. Then, we have
    \begin{equation}
        \Big|\sum_{x\in \mathbb{F}_{p^n}}\chi(f(x))\Big|\leq (l-1)p^{\frac{n}{2}}.\nonumber 
    \end{equation}
\end{lemma}

\section {Value Distribution of $S(a,b)$}
\label{Sec_ValDis_Sab}

\subsection{Evaluation of $S(a,b)$}

In this paper, the value distribution of $S(a,b)$ in (\ref{Def_Exp_Sum}) will be calculated as $a$ and $b$ run through $\mathbb{F}_{p^n}$ for the following parameters:

	\indent \hspace{5pt}$\bullet$ $p$ is an odd prime such that $p\equiv 3\bmod 4$;\\
	 \indent \hspace{5pt}$\bullet$ $n$ is an odd integer;\\
    \indent \hspace{5pt}$\bullet$ $d=(p^n+1)/(p^k+1)+(p^n-1)/2$ with $k|n$.

When either $a$ or $b$ is equal to zero, $S(a,b)$ is determined as in the following lemma.  
\begin{lemma}\label{Lem_Eval_Zero}
{
When either $a$ or $b$ is equal to zero, $S(a,b)$ is determined as
\begin{align*}
	S(a,b)=
	\begin{cases}
	p^n,&~{\rm when~}a=b=0\\
	0,&~{\rm when~}a\neq 0~{\rm and}~b=0\\
	\pm jp^{\frac{n}{2}},&~{\rm when~}a=0~{\rm and}~b\neq 0.
	\end{cases}
\end{align*}
}
\end{lemma}
\begin{proof}
The case of $b=0$ is easily proved. We need to prove the case when $a=0$ and $b\neq 0$. 
 Since $\gcd(d,p^n-1)=2$, we have
 \begin{align*}
 	S(0,b)=\sum_{x\in \mathbb{F}_{p^n}} \chi(bx^d)=\sum_{x\in \mathbb{F}_{p^n}} \chi(bx^2),~b\in \mathbb{F}_{p^n}^*.
 \end{align*}
 Note that ${\rm tr}_1^n(bx^2)$ is a quadratic form in $n$ variables over $\mathbb{F}_p$. From Lemma \ref{Lem_Cal_Rank}, it is straightforward that the quadratic form ${\rm tr}_1^n(bx^2)$ always has rank $n$. Hence, from Lemma \ref{Lema_QfEvaluation}, we have $S(0,b)=\pm jp^{\frac{n}{2}}$. \hfill $\Box$
\end{proof}

\vspace{0.1cm}

Next, we will calculate $S(a,b)$ for $a,b\in \mathbb{F}_{p^n}^*$. Note that $\gcd(p^k+1,p^n-1)=2$, $d(p^k+1)\equiv 2\bmod p^n-1$, and $-1$ is a nonsquare in $\mathbb{F}_{p^n}$. 
Replacing $x$ by $x^{p^k+1}$ for squares in $\mathbb{F}_{p^n}$ and $-x^{p^k+1}$ for nonsquares in $\mathbb{F}_{p^n}$, $S(a,b)$ is expressed in terms of quadratic forms as
\begin{align}\label{Eq_Def_ExpSum}
	S(a,b)&=\sum_{x\in \mathbb{F}_{p^n}}\chi (ax+bx^d)\nonumber\\
	&=\frac{1}{2}\Big(\sum_{x\in \mathbb{F}_{p^n}}\chi (ax^{p^k+1}+bx^2)+\sum_{x\in \mathbb{F}_{p^n}}\chi (-ax^{p^k+1}+bx^2)\Big)\nonumber\\
	&=\frac{1}{2}(S_1(a,b)+S_2(a,b))
\end{align}
where $S_1(a,b)=\sum_{x\in \mathbb{F}_{p^n}}\chi (ax^{p^k+1}+bx^2)$ and $S_2(a,b)=\sum_{x\in \mathbb{F}_{p^n}}\chi (-ax^{p^k+1}+bx^2)$. From Lemma \ref{Lem_Cor_Qf}, both
\begin{align*}
	Q_1(x)={\rm tr}_k^n(ax^{p^k+1}+bx^2)
\end{align*}
and
\begin{align*}
	Q_2(x)={\rm tr}_k^n(-ax^{p^k+1}+bx^2)
\end{align*}
are quadratic forms in $e$ variables over $\mathbb{F}_{p^k}$, where $e=n/k$. Thus, from Lemma \ref{Lema_QfEvaluation}, $S_1(a,b)$ and $S_2(a,b)$ can be computed if their ranks are obtained. From Lemma \ref{Lem_Cal_Rank}, in order to derive the rank of the quadratic form $Q_1(x)$, we need to count the number of solutions $x\in \mathbb{F}_{p^n}$ satisfying
\begin{align*}
	Q_1(x+y)-Q_1(x)-Q_1(y)=0,~{\rm for~all}~y\in \mathbb{F}_{p^n},
\end{align*}
which can be rewritten as
\begin{align*}
	\phi_{a,b}(x)=a^{p^k}x^{p^{2k}}+2b^{p^k}x^{p^k}+ax=0.
\end{align*}
Since the polynomial $\phi_{a,b}(x)$ is a linearized polynomial over $\mathbb{F}_{p^n}$ and its degree is $p^{2k}$, the number of roots $x\in \mathbb{F}_{p^n}$ of $\phi_{a,b}(x)$ can be $1$, $p^k$, or $p^{2k}$. Thus, from Lemma \ref{Lem_Cal_Rank}, $Q_1(x)$ can have the rank $e$, $e-1$, or $e-2$. Similarly, the corresponding linearized polynomial of $Q_2(x)$ is given as $\phi_{-a,b}(x)$ and the possible rank of $Q_2(x)$ is also $e$, $e-1$, or $e-2$. 

Therefore, from Lemma \ref{Lema_QfEvaluation}, each of $S_1(a,b)$ and $S_2(a,b)$ has the values
\begin{align}\label{Eq_Qf_Values}
	\begin{cases}
		\pm j p^{\frac{n}{2}},&~{\rm for~}r=e\\
		\pm \sqrt{p^k}p^{\frac{n}{2}},&~{\rm for~}r=e-1\\
		\pm j p^kp^{\frac{n}{2}},&~{\rm for~}r=e-2
	\end{cases}
\end{align}
where $r$ denotes the rank of the corresponding quadratic form.

However, there exist the values of $S(a,b)$ which actually do not occur when $a$ and $b$ run through $\mathbb{F}_{p^n}$ and they will be ruled out as in the following lemmas.

\vspace{0.1cm}

\begin{lemma}\label{Lem_OneRank}
	At least one of $\phi_{a,b}(x)$ and $\phi_{-a,b}(x)$ has a single root $x=0$ in $\mathbb{F}_{p^n}$, i.e., at least one of $Q_1(x)$ and $Q_2(x)$ always has the rank $e$.
\end{lemma}
\begin{proof} Assume that both $\phi_{a,b}(x)$ and $\phi_{-a,b}(x)$ have nonzero roots $x_1\in \mathbb{F}_{p^n}^*$ and $x_2\in \mathbb{F}_{p^n}^*$, respectively. Then, we have
\begin{align}\label{Eq_TwoEq}
	\phi_{a,b}(x_1)=0 \Leftrightarrow a^{p^k}x_1^{p^{2k}-1}+2b^{p^k}x_1^{p^k-1}+a&=0\nonumber\\
	\phi_{-a,b}(x_2)=0 \Leftrightarrow a^{p^k}x_2^{p^{2k}-1}-2b^{p^k}x_2^{p^k-1}+a&=0.
\end{align}
Using (\ref{Eq_TwoEq}), we can remove $2b^{p^k}$ as
\begin{align}\label{Eq_0508_001}
	a^{p^k}(x_1^{p^{2k}-p^k}+x_2^{p^{2k}-p^k})+a(x_1^{1-p^k}+x_2^{1-p^k})=0.
\end{align}
Since $x_1^{1-p^k}+x_2^{1-p^k}\neq 0$, (\ref{Eq_0508_001}) is rewritten as
\begin{align}\label{Eq2_Lem1}
	&~~~~a^{p^k-1}\frac{x_1^{p^{2k}-p^k}+x_2^{p^{2k}-p^k}}{x_1^{1-p^k}+x_2^{1-p^k}}=a^{p^k-1}(x_1x_2)^{p^k-1}(x_1+x_2)^{p^k-1}=-1.
\end{align}
The left-hand side of (\ref{Eq2_Lem1}) is the  $(p^k-1)$-th power in $\mathbb{F}_{p^n}$ while $-1$ is a nonsquare in $\mathbb{F}_{p^n}$, which is a contradiction. Hence the proof is done. \hfill $\Box$
\end{proof}

In \cite{Xia}, they used the wise method to exclude some redundant values of $S(a,b)$ by using the Weil's bound in Lemma \ref{Thrm_Weil}. Similarly, the following lemma is stated. 
\vspace{0.1cm}
\begin{lemma}\label{Exc_Weil}
	Two candidate values of $S(a,b)$, $\pm j(p^k-1)/2p^{n/2}$,
	 do not actually occur when $a$ and $b$ run through $\mathbb{F}_{p^n}^*$. 
\end{lemma}
\begin{proof}
	If the rank of $Q_2(x)$ is odd, $S_2(a,b)$ has a pure imaginary value in (\ref{Eq_ExpValues}). Then we have
	\begin{align*}
		S_1(a,b)&=2\sum_{x\in C_0} \chi(ax^{\frac{p^k+1}{2}}+bx)+1\\
		-S_2(a,b)&=\sum_{x\in \mathbb{F}_{p^n}}\chi (ax^{p^k+1}-bx^2)=2\sum_{x\in C_1} \chi(ax^{\frac{p^k+1}{2}}+bx)+1
	\end{align*}
where $C_0$ and $C_1$ are sets of squares and nonsquares in $\mathbb{F}_{p^n}^*$, respectively. Hence we have
	\begin{align*}
		\sum_{x\in \mathbb{F}_{p^n}}\chi(ax^{\frac{p^k+1}{2}}+bx)=\frac{1}{2}(S_1(a,b)-S_2(a,b)).
	\end{align*} 
	Assume that $S_1(a,b)=\pm jp^{n/2}$ and $S_2(a,b)=\mp jp^kp^{n/2}$ or vice versa. Then we have
	\begin{align*}
		\Big|\sum_{x\in \mathbb{F}_{p^n}}\chi(ax^{\frac{p^k+1}{2}}+bx)\Big|=\frac{p^k+1}{2}p^{\frac{n}{2}},
	\end{align*}
	which contradicts the Weil bound in Lemma \ref{Thrm_Weil}. Thus the values   $S(a,b)=(S_1(a,b)+S_2(a,b))/2=\pm j(p^k-1)p^{n/2}/2$ are excluded.  \hfill $\Box$
\end{proof}

Using the above lemmas, the possible candidate values of $S(a,b)$ can be derived as in the following theorem. 

\begin{theorem}\label{Thm_Eval_Exp}
	$S(a,b)$ for $a,b\in \mathbb{F}_{p^n}$ has the following candidate values 
	\begin{align}\label{Eq_ExpValues}
		\{p^n,0,\pm jp^{\frac{n}{2}}, \frac{\sqrt{p^k}\pm j}{2}p^{\frac{n}{2}}, \frac{-\sqrt{p^k}\pm j}{2}p^{\frac{n}{2}}, \pm j\frac{p^k+1}{2}p^{\frac{n}{2}}\}.
	\end{align}
\end{theorem}
\begin{proof} From Lemmas  \ref{Lem_Eval_Zero}--\ref{Exc_Weil} and (\ref{Eq_Qf_Values}), the proof is easily done.  \hfill $\Box$
\end{proof}

The above theorem also indicates that the magnitudes of the cross-correlation values of a $p$-ary m-sequence and its decimated sequences by $d$ are upper bounded by $\sqrt{1+\big((p^k+1)/2\big)^2 p^{n}}\approx (p^k+1)\sqrt{N}/2$.

\subsection{Value Distribution of $S(a,b)$}

Using the result in Theorem \ref{Thm_Eval_Exp}, we will derive the value distribution of $S(a,b)$. Let $v_i,~0\leq i\leq 9$, be the $i$-th value in (\ref{Eq_ExpValues}), that is, $v_0=p^n$, $v_1=0$, $v_2=jp^{\frac{n}{2}}$, $v_3=-v_2$, $v_4=\frac{\sqrt{p^k}+ j}{2}p^{\frac{n}{2}}$, $v_5=v_4^*$, $v_6=\frac{-\sqrt{p^k}+ j}{2}p^{\frac{n}{2}}$, $v_7=v_6^*$, $v_8=j\frac{p^k+1}{2}p^{\frac{n}{2}}$, and $v_9=-v_8$. Let $\Omega_i,~0\leq i\leq 9$, be the number of occurrences of $v_i$ when $a$ and $b$ run through $\mathbb{F}_{p^n}$ and clearly, $\Omega_0=1$. Since $S(-a,-b)=S^*(a,b)$,  each conjugate pair in (\ref{Eq_ExpValues}) has the same number of occurrences, that is, $\Omega_2=\Omega_3$, $\Omega_4=\Omega_5$, $\Omega_6=\Omega_7$, and $\Omega_8=\Omega_9$.  Hence we need five independent equations in terms of $\Omega_i$'s to determine the entire value distribution. Since $\sum_{x\in \mathbb{F}_{p^n}}\chi(ax)=0$ for any $a\in \mathbb{F}_{p^n}^*$, it is straightforward to obtain the following three equations
\begin{align}
	\sum_{i=0}^9\Omega_i&=p^{2n}\label{Eq_0507_01}\\
	\sum_{i=0}^9 v_i\Omega_i&=\sum_{a,b\in \mathbb{F}_{p^n}}S(a,b)\nonumber\\&=\sum_{b,x\in \mathbb{F}_{p^n}}\chi(bx^d)\sum_{a\in \mathbb{F}_{p^n}}\chi(ax)=p^{2n}\label{Eq_0507_02}
\end{align}
and 
\begin{align}
	\sum_{i=0}^9 v_i^2\Omega_i&=\sum_{a,b\in \mathbb{F}_{p^n}}S^2(a,b)\nonumber\\&=\sum_{b,x,y\in \mathbb{F}_{p^n}}\chi(b(x^d+y^d))\sum_{a\in \mathbb{F}_{p^n}}\chi(a(x+y))\nonumber\\&=p^n\sum_{b,y\in \mathbb{F}_{p^n}} \chi(2by^d)=p^{2n}.\label{Eq_0507_03}
\end{align}

\vspace{0.1cm}

\begin{lemma}[Theorems 4.6, 5.4, and 5.6 \cite{Bluher}]
\label{Lemma_Zeng}
{
Let $f(z)=z^{p^s+1}-\psi z+\psi$, $\psi\in \mathbb{F}_{p^n}^*$. Then $f(z)$ has either 0, 1, 2, or $p^{\gcd(s,n)}+1$ roots in $\mathbb{F}_{p^n}^*$. The number of $\psi\in \mathbb{F}_{p^n}^*$ such that $f(z)$
 has exactly one root in $\mathbb{F}_{p^n}^*$ is equal to $p^{n-\gcd(s,n)}$. If $z_0\in \mathbb{F}_{p^n}^*$ is the unique root of the equation, then $z_0$ should satisfy
\begin{align}\label{0518_A}
  (z_0-1)^{\frac{p^n-1}{p^{\gcd(s,n)}-1}}=1.
  \end{align}
    The number of $\psi\in \mathbb{F}_{p^n}^*$ such that $f(z)$
 has exactly $p^{\gcd(s,n)}+1$ roots in $\mathbb{F}_{p^n}^*$ is equal to $\frac{p^{n-\gcd(s,n)}-1}{p^{2\gcd(s,n)}-1}$. Any root $z_0\in \mathbb{F}_{p^n}^*$ from the $p^{\gcd(s,n)}+1$ roots should satisfy (\ref{0518_A}). 
} 
\end{lemma}
\vspace{0.1cm}

From the above lemma, the remaining two equations in terms of $\Omega_i$'s are obtained as in the following lemma.

\vspace{0.1cm}
 
\begin{lemma} We have
	\begin{align}
		N_1&=\Omega_4+\Omega_5+\Omega_6+\Omega_7=2p^{n-k}(p^n-1)\label{Eq_0507_005}\\
		N_2&=\Omega_8+\Omega_9=\frac{2(p^{n-k}-1)(p^n-1)}{p^{2k}-1}.\label{Eq_0507_006}
	\end{align}
\end{lemma}
\begin{proof}
From (\ref{Eq_Qf_Values}), Lemmas \ref{Lem_Cal_Rank} and \ref{Lem_OneRank}, $N_1$ is equal to the number of $(a,b)\in \mathbb{F}_{p^n}^*\times \mathbb{F}_{p^n}^*$ such that either $\phi_{a,b}(x)$ or $\phi_{-a,b}(x)$ in (\ref{Eq_TwoEq}) has $p^k$ roots in $\mathbb{F}_{p^n}$. Similarly, $N_2$ is equal to the number of $(a,b)\in \mathbb{F}_{p^n}^*\times \mathbb{F}_{p^n}^*$ such that either $\phi_{a,b}(x)$ or $\phi_{-a,b}(x)$ has $p^{2k}$ roots in $\mathbb{F}_{p^n}$. 
	
	Consider $\phi_{a,b}(x)$ and let $x^{p^k-1}=y$. Then $\phi_{a,b}(x)/x$ in (\ref{Eq_TwoEq}) is written as
	\begin{align}\label{0515_A}
		a^{p^k}y^{p^{k}+1}+2b^{p^k}y+a=0.
	\end{align}
	Let $y_1$ and $y_2$ be the distinct solutions to (\ref{0515_A}). Then we have
	\begin{align*}
		y_1y_2(y_1-y_2)^{p^k}&=y_1^{p^k+1}y_2-y_1y_2^{p^k+1}\\
		&=-y_2\Big(\frac{2b^{p^k}y_1+a}{a^{p^k}}\Big)+y_1\Big(\frac{2b^{p^k}y_2+a}{a^{p^k}}\Big)\\
		&=\frac{y_1-y_2}{a^{p^k-1}}.
	\end{align*}
	Thus we have
	\begin{align*}
		y_1y_2=(y_1-y_2)^{1-p^k}a^{1-p^k},
	\end{align*}
	which means that both $y_1$ and $y_2$ are $(p^k-1)$-th power in $\mathbb{F}_{p^n}$ or both $y_1$ and $y_2$ are not $(p^k-1)$-th power in $\mathbb{F}_{p^n}$. Therefore, letting  $z=(-2b^{p^k}/a)y=-(2b^{p^k}/a)x^{p^k-1}$, we have
	\begin{align}
	&~~~~~\phi_{a,b}(x)~{\rm has}~ p^k-1 ~{\rm nonzero~ roots~in}~ \mathbb{F}_{p^n}\nonumber\\
	 &\Leftrightarrow (\ref{0515_A})~{\rm has~ a~single~ solution}~y_0 \in \mathbb{F}_{p^n}^*{~\rm and~}\frac{az_0}{2b^{p^k}}=-\zeta^{p^k-1}~{\rm for~some~}\zeta\in \mathbb{F}_{p^n}^*\label{Eq_0507_010}
\end{align}
and
\begin{align}
	&~~~~~\phi_{a,b}(x)~{\rm has}~ p^{2k}-1 ~{\rm nonzero~ roots~in}~ \mathbb{F}_{p^n}\nonumber\\
	 &\Leftrightarrow (\ref{0515_A})~{\rm has}~ p^k+1~ {\rm nonzero~ solutions~}y_0\in\mathbb{F}_{p^n}^*{~\rm and~any~nonzero~solution}~\nonumber\\&~~~~{\rm from~the~}p^k+1~{\rm solutions~satisfies~that~}\frac{az_0}{2b^{p^k}}=-\zeta^{p^k-1}~{\rm for~some~}\zeta\in \mathbb{F}_{p^n}^*\label{Eq_0507_011}
\end{align}
where $y_0=(-a/(2b^{p^k}))z_0$.

	Let $\gamma=4b^{p^{2k}+p^k}/a^{2p^k}$. Then $\phi_{a,b}(x)$ in (\ref{Eq_TwoEq}) is rewritten as 
	\begin{align}\label{Poly_z}
		z^{p^k+1}-\gamma z+\gamma=0,
	\end{align}
which has the same form as the polynomial in Lemma \ref{Lemma_Zeng}. From Lemma \ref{Lemma_Zeng}, when the number of roots $z\in \mathbb{F}_{p^n}^*$ of (\ref{Poly_z}) is $1$ or $p^{k}+1$, $\psi$ is always a square in $\mathbb{F}_{p^n}$ because 
\begin{align}\label{Eq_0507_030}
	(z_0-1)^{\frac{p^n-1}{p^k-1}}=\Big(\frac{z_0^{p^k+1}}{\psi}\Big)^{\frac{p^n-1}{p^k-1}}=1,
\end{align}
where $z_0$ is a root of $f(z)$ and $(p^n-1)/(p^k-1)$ is odd. Fortunately, $\gamma=4b^{p^{2k}+p^k}/a^{2p^k}$ is a square in $\mathbb{F}_{p^n}$ in this case. Thus, Lemma \ref{Lemma_Zeng} can be used for the proof of this theorem.

Now, we will calculate the number of $(a,b)\in \mathbb{F}_{p^n}^* \times \mathbb{F}_{p^n}^*$ satisfying (\ref{Eq_0507_010}) and (\ref{Eq_0507_011}), respectively. Lemma \ref{Lemma_Zeng} tells us that the number of $\gamma\in \mathbb{F}_{p^n}^*$ such that (\ref{Poly_z}) has a single solution in $\mathbb{F}_{p^n}^*$ is $p^{n-k}$ and the number of $\gamma\in \mathbb{F}_{p^n}^*$ such that (\ref{Poly_z}) has $p^k+1$ solutions in $\mathbb{F}_{p^n}^*$ is $(p^{n-k}-1)/(p^{2k}-1)$. Since for any $b\in \mathbb{F}_{p^n}^*$, $\gamma$ runs through squares in $\mathbb{F}_{p^n}^*$ twice as $a$ runs through $\mathbb{F}_{p^n}^*$, the number of $(a,b)$ such that (\ref{Poly_z}) has a single solution is $2p^{n-k}(p^n-1)$ and the number of $(a,b)$ such that (\ref{Poly_z}) has $p^k+1$ solutions is $2(p^{n-k}-1)(p^n-1)/(p^{2k}-1)$.

Now, from (\ref{Eq_0507_010}) and (\ref{Eq_0507_011}), we have to check that each solution $z_0$ of (\ref{Poly_z}) satisfies ${az_0}/({2b^{p^k}})=-\zeta^{p^k-1}$ for some $\zeta\in \mathbb{F}_{p^n}^*$. Substituting  $\gamma=4b^{p^{2k}+p^k}/a^{2p^k}$, (\ref{Eq_0507_030}) is rewritten as
\begin{align}
	(z_0-1)^{\frac{p^n-1}{p^k-1}}=\Big(\big(\frac{az_0}{2b^{p^k}}\big)^{p^k+1}\Big)^{\frac{p^n-1}{p^k-1}}=\Big(\frac{az_0}{2b^{p^k}}\Big)^{\frac{2(p^n-1)}{p^k-1}}=1.\label{Eq_0507_031}
\end{align}
Note that $a=\pm \mu \in \mathbb{F}_{p^n}^*$ map to the same value of $\gamma$. From (\ref{Eq_0507_031}), we have
\begin{align}\label{0515_005}
	\frac{az_0}{2b^{p^k}}=\rho^{\frac{p^k-1}{2}}
\end{align}
for some $\rho\in \mathbb{F}_{p^n}^*$. Then, in order to satisfy (\ref{Eq_0507_010}) and (\ref{Eq_0507_011}), we have
\begin{align}\label{0518_B}
	-\rho^{\frac{p^k-1}{2}}=\alpha^{\frac{p^k-1}{2}\cdot\frac{p^n-1}{p^k-1}}\rho^{\frac{p^k-1}{2}}= \zeta^{{p^k-1}}.
\end{align}
From (\ref{0518_B}), $\rho$ must be a nonsquare in $\mathbb{F}_{p^n}$. However, in (\ref{0515_005}), one of $a=\pm \mu \in \mathbb{F}_{p^n}^*$ gives a square $\rho$ and the other gives a nonsquare $\rho$. Thus, one of $a=\pm \mu \in \mathbb{F}_{p^n}^*$ should be excluded from the counting of  $(a,b)$. Hence, the number of $(a,b)$ satisfying (\ref{Eq_0507_010}) and (\ref{Eq_0507_011}) is $p^{n-k}(p^n-1)$ and $(p^{n-k}-1)(p^n-1)/(p^{2k}-1)$, respectively.  

For the case of $\phi_{-a,b}(x)$, we just consider $-a$ instead of $a$. Similarly to the previous case, the number of $(a,b)$ such that $\phi_{-a,b}(x)$ has $p^k$ roots is $p^{n-k}(p^n-1)$ and the number of $(a,b)$ such that $\phi_{-a,b}(x)$ has $p^{2k}$ roots is $(p^{n-k}-1)(p^n-1)/(p^{2k}-1)$.

From Lemma \ref{Lem_OneRank}, one of $\phi_{a,b}(x)$ and $\phi_{-a,b}(x)$ always has a single root in $\mathbb{F}_{p^n}$. Therefore, there exist no intersection of $(a,b)$ between the previous two cases, that is, 
\begin{align*}
	N_1&=2p^{n-k}(p^n-1)\\
	N_2&=\frac{2(p^{n-k}-1)(p^n-1)}{{p^{2k}-1}}. 
\end{align*} \hfill $\Box$
\end{proof}

Thus, the value distribution of $S(a,b)$ can be derived as follows.

\begin{theorem}\label{Value_Dist}
As $a$ and $b$ run through $\mathbb{F}_{p^n}$, the value distribution of $S(a,b)$ is determined as 
  \begin{align*}
        S(a,b)=
        \begin{cases}
        	    p^n,&{\rm once}\\
            0,&\frac{(p^k-1)(p^{2n}-1)}{2(p^k+1)}~{\rm times}\\
            \pm jp^{n/2},&\frac{p^{2n}-1}{4}-\frac{(p^n-1)^2}{2(p^k-1)}~{\rm times}\\
           \frac{\sqrt{p^k}\pm j}{2}p^{\frac{n}{2}},&\frac{(p^n-1)(p^{n-k}+p^{\frac{n-k}{2}})}{2}~{\rm times}\\
            \frac{-\sqrt{p^k}\pm j}{2},&\frac{(p^n-1)(p^{n-k}-p^{\frac{n-k}{2}})}{2}~{\rm times}\\
            \pm j\frac{p^k+1}{2}p^{\frac{n}{2}},&\frac{(p^{n-k}-1)(p^n-1)}{p^{2k}-1}~{\rm times}.
        \end{cases}
     \end{align*}
\end{theorem}
\begin{proof}
	Clearly, $S(a,b)=p^n$ occurs once when $a=b=0$. The five independent equations have already been derived as
	\begin{align*}
		&\sum_{i=1}^9\Omega_i=p^{2n}-1\\
		&\sum_{i=1}^9 v_i\Omega_i=p^{2n}-p^n\\
		&\sum_{i=1}^9 v_i^2\Omega_i=0\\ 
		&\Omega_4+\Omega_5+\Omega_6+\Omega_7=2p^{n-k}(p^n-1)\\
		&\Omega_8+\Omega_9=\frac{2(p^{n-k}-1)(p^n-1)}{p^{2k}-1}.
	\end{align*}
	Solving the above five equations, we can prove the theorem.   \hfill $\Box$
\end{proof}

\section{Sequence Family $\mathcal{G}$}
\label{Sec_Seq_Fam}

Each sequence with period $N=p^n-1$ in the sequence family $\mathcal{G}$ is defined as
\begin{align*}
	s_{\beta}(t)={\rm tr}_1^n(\alpha^t)+{\rm tr}_1^n(\beta \alpha^{dt}),~0\leq t\leq N-1
\end{align*}
where $\beta\in \mathbb{F}_{p^n}$. For $\beta_1,\beta_2\in \mathbb{F}_{p^n}$, the correlation function between the sequences $s_{\beta_1}(t)$ and $s_{\beta_2}(t)$ in $\mathcal{G}$ at shift value $\tau$ is given as
\begin{align}\label{Cor_Seq_Fam}
	C_{s_{\beta_1},s_{\beta_2}}(\tau)&=\sum_{t=0}^{N-1} \omega^{s_{\beta_1}(t+\tau)+s_{\beta_2}(t)}\\&=-1+\sum_{x\in \mathbb{F}_{p^n}} \chi\big((\delta-1)
	x+(\beta_1 \delta^d-\beta_2)x^d\big)\\&=-1+S(\delta-1,\beta_1\delta^d-\beta_2)
\end{align}
where $\delta=\alpha^{\tau}$ and $x=\alpha^t$. Thus, the correlation function $C_{s_{\beta_1},s_{\beta_2}}(\tau)$ is expressed in terms of $S(a,b)$. From Theorem \ref{Value_Dist}, the upper bound of correlation values of $p$-ary sequences in $\mathcal{G}$ is easy to derive as in the following theorem.

\begin{theorem}
The family size of $\mathcal{G}$ is $p^n$ and the magnitudes of the correlation values of sequences in $\mathcal{G}$ are upper bounded by
\begin{align*}
	|C_{s_{i},s_{j}}(\tau)|\leq \sqrt{1+\big((p^k+1)/2\big)^2 p^{n}}\approx \frac{p^k+1}{2} \sqrt{N}, ~i\neq j~{\rm or~}i=j, \tau\neq 0 
\end{align*}
where $s_i,s_j\in \mathcal{G}$ and $\tau$ is the shift value. 
\end{theorem}

\section{The Weight Distribution of $\mathcal {C}$}
\label{WeightDist_C}

Let $\mathcal {C}$ be the cyclic code over $\mathbb{F}_p$ with length $N=p^n-1$ in which each codeword is defined as
\begin{align*}
	c(a,b)=(c_0,c_1,\dots,c_{N-1}),~a,b\in \mathbb{F}_{p^n}
\end{align*} 
where $c_i={\rm tr}_1^n (a\alpha^i+b\alpha^{di})$, $0\leq i \leq N-1$. The Hamming weight of the codeword $c(a,b)$ is expressed as
\begin{align}\label{Eq_WD}
	H_w(c(a,b))&=\big|\{i|0\leq i\leq N-1,c_i\neq 0\}\big|\nonumber\\
	&=N-\big|\{i|0\leq i\leq N-1,c_i= 0\}\big|\nonumber\\
	&=N-\frac{1}{p}\sum_{i=0}^{N-1}\sum_{l=0}^{p-1}(\chi(a\alpha^i+b\alpha^{di}))^l\nonumber\\
	&=N-\frac{N}{p}+\frac{p-1}{p}-\frac{1}{p}\sum_{l=1}^{p-1}S(la,lb)\nonumber\\
	&=p^{n-1}(p-1)-\frac{1}{p}\sum_{l=1}^{p-1}S(la,lb)\nonumber\\
	&=p^{n-1}(p-1)-\frac{1}{p}\mu(S(a,b))
\end{align}
where $\mu(S(a,b))=\sum_{l=1}^{p-1}S(la,lb)$. Hence, the Hamming weight of the codeword $c(a,b)$ is determined by calculating $\mu(S(a,b))$ for each value of $S(a,b)$. Let $\{w_0,w_1,\dots,w_N\}$ be the weight distribution of $\mathcal{C}$, where $w_i$ is the number of occurrences of the codewords $c(a,b)$ of Hamming weight $i,~0\leq i \leq N$, as $a$ and $b$ run through $\mathbb{F}_{p^n}$. The following lemma is needed for the calculation of $\mu(S(a,b))$. 
\begin{lemma}[Lemma 4 in \cite{Feng}]\label{Feng_AM}
Let $\omega$ be a primitive $p$-th root of unity and $(\frac{\cdot}{p})$ the Legendre symbol. The Galois group of $\mathbb{Q}(\omega)$ over $\mathbb{Q}$ is $\{\sigma_i|1\leq i\leq p-1\}$,  where the automorphism $\sigma_i$ of $\mathbb{Q}(\omega)$ is determined by $\sigma_i(\omega)=\omega^i$. The unique quadratic subfield of $\mathbb{Q}(\omega)$ is $\mathbb{Q}(\sqrt{p^*})$, where $p^*=(\frac{-1}{p})p$ and $\sigma_i(\sqrt{p^*})=(\frac{i}{p})\sqrt{p^*},~1\leq i\leq p-1$. 
\end{lemma}

Using (\ref{Eq_WD}) and Lemma \ref{Feng_AM}, the weight distribution of the cyclic code is given as follows. 

\begin{theorem}\label{Thm_WeightDist}
	The weight distribution $\{w_0,w_1,\dots,w_N\}$ of the cyclic code $\mathcal{C}$ over $\mathbb{F}_p$ with the length $N$ and the dimension ${\rm dim}_{\mathbb{F}_p}\mathcal{C}=2n$ is given as
	\begin{align*}
	w_i=
	\begin{cases}
	1,&~{\rm when~}i=0\\
	{(p^n-1)(p^{n}-2p^{n-k}+1)},&~{\rm when~}i= p^{n-1}(p-1)\\
	(p^n-1)(p^{n-k}-p^{\frac{n-k}{2}}),&~{\rm when~}i= (p-1)(p^{n-1}+\frac{1}{2}p^{\frac{n+k}{2}-1})\\
	(p^n-1)(p^{n-k}+p^{\frac{n-k}{2}}),&~{\rm when~}i= (p-1)(p^{n-1}-\frac{1}{2}p^{\frac{n+k}{2}-1}).
	\end{cases}
	\end{align*}
\end{theorem}
\begin{proof} From (\ref{Eq_WD}), we have to calculate 
\begin{align*}
	\mu (S(a,b))=\sum_{l=1}^{p-1}S(la,lb)=\sum_{l=1}^{p-1}\sigma_l(S(a,b))
\end{align*}
to determine $H_w(c(a,b))$. Since $p\equiv 3\bmod 4$ and $n$ is odd, $\pm jp^{\frac{n}{2}}$ is equal to $\pm (\sqrt{p^*})^n$. Hence, from Lemma \ref{Feng_AM}, we can determine the image of $\mu$ map of each value of $S(a,b)$ as
\begin{align*}
	&\mu(0)=\mu(\pm jp^{\frac{n}{2}})=\mu(\pm j\frac{p^k+1}{2}p^{\frac{n}{2}})\\
	&~~~~~~=\sum_{l=1}^{p-1}\sigma_l(\pm \sqrt{p^*})^n=(\pm \sqrt{p^*})^n\sum_{l=1}^{p-1}(\frac{l}{p})=0\\
	&\mu(\frac{\sqrt{p^k}}{2}p^{\frac{n}{2}} \pm\frac{1}{2}jp^{\frac{n}{2}})=\frac{\sqrt{p^k}}{2}(p-1)p^{\frac{n}{2}}\\
	&\mu(-\frac{\sqrt{p^k}}{2}p^{\frac{n}{2}} \pm\frac{1}{2}jp^{\frac{n}{2}})=-\frac{\sqrt{p^k}}{2}(p-1)p^{\frac{n}{2}}\\
	&\mu(p^n)=(p-1)p^n.
\end{align*}
Thus, from (\ref{Eq_WD}), the proof is done.   \hfill $\Box$
\end{proof}

\section{Conclusion}
\label{Sec_Con}

In this paper, the value distribution of the exponential sum $S(a,b)$ as $a$ and $b$ run through $\mathbb{F}_{p^n}$ is derived. Using the result, we construct the sequence family $\mathcal{G}$ in which each sequence has the period of $N=p^n-1$. The family size is $p^n$ and the correlation magnitude is roughly upper bounded by $(p^k+1)\sqrt{N}/2$. The weight distribution of the cyclic code $\mathcal{C}$ over $\mathbb{F}_p$ with the length $N$ and the dimension ${\rm dim}_{\mathbb{F}_p}\mathcal{C}=2n$ is also determined. Our result includes the result in \cite{Xia} as a special case.




\begin{thebibliography}{99}
%
%
\bibitem{Trachtenberg} H. M. Trachtenberg, ``On the cross-correlation
functions of maximal recurring sequences," {\em Ph.D. dissertation},
Univ. of Southern California, Los Angeles, CA, 1970.

\bibitem{Helleseth} T. Helleseth, ``Some results about the
cross-correlation function between two maximal linear
sequences," {\em Discrete Math.}, vol. 16, pp. 209--232, 1976.

\bibitem{Xia} Y. Xia, X. Zeng, and L. Hu, ``Further crosscorrelation properties of sequences with the decimation factor $d=(p^n+1)/(p+1)-(p^n-1)/2$," {\em Appl. Algebra Eng. Commun. Comput.},  vol. 21, no. 5, pp. 329--342, 2010. 

\bibitem{Vlugt} M. Van Der Vlugt, ``Surfaces and the weight distribution of a family of codes," {\em IEEE Trans. Inf. Theory}, vol. 43, no. 4, pp. 1354--1360, Jul. 1997.

\bibitem{Yuan} J. Yuan, C. Carlet, and C. Ding, ``The weight distribution of a class of linear codes from perfect nonlinear functions," {\em IEEE Trans. Inf. Theory}, vol. 52, no. 2, pp. 712--717, Feb. 2006.

\bibitem{Feng2} K. Feng and J. Luo, ``Value distributions of exponential sums from perfect nonlinear functions and their applications," {\em IEEE Trans. Inf. Theory}, vol. 53, no. 9, pp. 3035--3041, Sep. 2007.

\bibitem{Gold} R. Gold, ``Maximal recursive sequences with 3-valued recursive cross-correlation functions," {\em IEEE Trans. Inf. Theory}, vol. 14,
no. 1, pp. 154--156, Jan. 1968.


\bibitem{Kumar} P. V. Kumar and O. Moreno,
``Prime-phase sequences with periodic correlation properties better than binary sequences," {\em IEEE Trans. Inf. Theory}, vol. 37, no. 3, pp.
603-616, May 1991.

\bibitem{Seo2} E.-Y. Seo, Y.-S. Kim, J.-S. No, and D.-J. Shin,
``Cross-correlation distribution of $p$-ary m-sequence and its $p+1$ decimated sequences with shorter period," {\em IEICE Trans. Fund. Electron., Commun. Comp. Sci.}, vol. E90-A,
no. 11, pp. 2568--2574, Nov. 2007.

\bibitem{Choi2} S.-T. Choi, T. Lim, and J.-S. No, ``On the cross-correlation of a $p$-ary m-sequence of period $p^{2m}-1$ and its decimated sequences by $(p^m+1)^2/(2(p+1))$," {\em IEEE Trans. Inf. Theory}, vol. 58, no. 3, pp. 1873--1879, Mar. 2012.

\bibitem{Kim} J.-Y. Kim, S.-T. Choi, J.-S. No, and H. Chung, ``A new family of $p$-ary sequences of period $(p^n-1)/2$ with low correlation," {\em IEEE Trans. Inf. Theory}, vol. 57, no. 6, pp. 3825--3830, Jun. 2011.

\bibitem{Kim2} D. S. Kim, H.-J. Chae, and H.-Y. Song, ``A generalization of the family of $p$-ary decimated sequences with low correlation," {\em IEEE Trans. Inf. Theory}, vol. 57, no. 11, pp. 7614--7617, Nov. 2011.

\bibitem{Feng} K. Feng and J. Luo, ``Weight distribution of some reducible cyclic codes," {\em Finite Fields Appl.}, vol. 14, no. 2, pp. 390--409, Apr. 2008.

\bibitem{Luo} J. Luo and K. Feng, ``Cyclic codes and sequences from generalized Coulter-Matthew function," {\em IEEE Trans. Inf. Theory}, vol. 54, no. 12, pp. 5345--5353, Dec. 2008.

\bibitem{Luo2} J. Luo and K. Feng, ``On the weight distributions of two classes of cyclic codes," {\em IEEE Trans. Inf. Theory}, vol. 54, no. 12, pp. 5332--5344, Dec. 2008.

\bibitem{Luo3} J. Luo, Y. Tang, and H. Wang, ``Cyclic codes and sequences: the generalized Kasami case," {\em IEEE Trans. Inf. Theory}, vol. 56, no. 5, pp. 2130--2142, May 2010.



\bibitem{Lidl} R.~Lidl and H.~Niederreiter, {\em Finite Fields},
vol. 20 of {\em Encyclopedia of Mathematics and Its Applications}.
Reading, MA: Addison-Wesley, 1983.

\bibitem{Bluher} A. W. Bluher, ``On $x^{q+1}+ax+b$," {\em Finite Fields Appl.}, vol. 10, no. 3, pp. 285-305, Jul. 2004.

\bibitem{Choi} S.-T. Choi and J.-S. No, ``On the cross-correlation distributions of $p$-ary m-sequences and their decimated sequences," accepted for publication in {\em IEICE Trans. Fund. Electron., Commun. Comp. Sci.}, Dec. 2011.

\bibitem{Choi3} S. T. Choi, T. Lim, J. S. No, and H. Chung, “Weight distribution of some cyclic codes,” to appear in {\em Proc. IEEE Int. Symp. Inf. Theory}, Cambridge, MA, USA, Jun.
2012.

\end{thebibliography}


\end{document}